# Features of the Earth's seasonal hydroclimate: Characterizations and comparisons across the Köppen-Geiger climates and across continents

Georgia Papacharalampous[1,*], Hristos Tyralis[2,3], Yannis Markonis[1], Petr Máca[1], and Martin Hanel[1]

1 Department of Water Resources and Environmental Modeling, Faculty of Environmental Sciences, Czech University of Life Sciences, Kamýcá 129, Praha-Suchdol 16500, Prague, Czech Republic

2 Department of Water Resources and Environmental Engineering, School of Civil Engineering, National Technical University of Athens, Heroon Polytechneiou 5, 15780 Zographou, Greece

3 Construction Agency, Hellenic Air Force, Mesogion Avenue 227–231, 15561 Cholargos, Greece

* Correspondence: papacharalampous.georgia@gmail.com, tel: +30 69474 98589

---

**Email addresses and ORCID profiles:** papacharalampous.georgia@gmail.com, https://orcid.org/0000-0001-5446-954X (Georgia Papacharalampous); montchrister@gmail.com, hristos@itia.ntua.gr, https://orcid.org/0000-0002-8932-4997 (Hristos Tyralis); markonis@fzp.czu.cz, https://orcid.org/0000-0003-0144-8969 (Yannis Markonis); maca@fzp.czu.cz, https://orcid.org/0000-0002-4972-3993 (Petr Máca); hanel@fzp.czu.cz, https://orcid.org/0000-0001-8317-6711 (Martin Hanel)

---

---

**Abstract:** Detailed investigations of time series features across climates, continents and variable types can progress our understanding and modelling ability of the Earth's hydroclimate and its dynamics. They can also improve our comprehension of the climate classification systems appearing in their core. Still, such investigations for seasonal hydroclimatic temporal dependence, variability and change are currently missing from the literature. Herein, we propose and apply at the global scale a methodological framework for filling this specific gap. We analyse over 13 000 earth-observed quarterly temperature, precipitation and river flow time series. We adopt the Köppen-Geiger climate classification system and define continental-scale geographical regions for conducting upon them seasonal hydroclimatic feature summaries. The analyses rely on three sample autocorrelation features, a temporal variation feature, a spectral entropy feature, a Hurst feature, a trend strength feature and a seasonality strength feature. We find notable differences to characterize the magnitudes of these features across the various Köppen-Geiger climate classes, as well as between continental-scale geographical regions. We, therefore, deem that the consideration of the comparative summaries could be beneficial in water resources engineering contexts. Lastly, we apply explainable machine learning to compare the investigated features with respect to how informative they are in distinguishing either the main Köppen-Geiger climates or the continental-scale regions. In this regard, the sample autocorrelation, temporal variation and seasonality strength features are found to be more informative than the spectral entropy, Hurst and trend strength features at the seasonal time scale.

**Keywords**: comparative hydrology; explainable machine learning; hydroclimatic features; Köppen-Geiger climate classification; precipitation; seasonality; streamflow; temperature; temporal dependence; trends

## 1. Introduction

The study of temporal and spatial aspects of the various hydroclimatic phenomena (e.g., the ones linked to temperature, precipitation or streamflow variables) holds a prominent position in Earth system science and engineering (see, e.g., the detailed lists of research topics compiled by Montanari et al. 2013 and Blöschl et al. 2019), with a large variety of hydroclimatic features being investigated with increasing frequency. Such investigations, as well as their underlying methodologies, are indeed necessary either when referring to high-impact case studies (i.e., case studies conducted for areas with large engineering interest; see, e.g., Montanari 2012), or when based on large-sample datasets (see, e.g., Archfield et al. 2014; Ledvinka and Lamacova 2015; Fischer and Schumann 2018; Hall and Blöschl 2018; Hanel et al. 2018; Markonis et al. 2018; Tyralis et al. 2018; Jehn et al. 2020; Messager et al. 2021; Papacharalampous et al. 2021; Slater et al. 2021), and often focus on hydroclimatic temporal dependence (which can be characterized by estimating the autocorrelation of hydroclimatic time series), variability and change. An overview of the related literature, summarizing information on both mean- and extreme-value hydroclimatic time series analyses, can be found in Papacharalampous et al. (2021).

Detailed comparisons of hydroclimatic features across different climate classes and geographical regions have their own key role towards improving our understanding and modelling ability of the Earth's hydroclimate and its dynamics (through the identification of climates and/or regions with similar hydroclimatic characteristics) and, at the same time, facilitate a direct connection with the climate literature. From a different point of view, they can also progress our comprehension of the climate classification systems appearing in their core, especially when conducted at the global scale (see, e.g., Beck et al. 2005; Markonis et al. 2018; Tyralis et al. 2018; Messager et al. 2021; Slater et al. 2021; Papacharalampous et al. 2022). Nonetheless, such comparisons in terms of hydroclimatic temporal dependence, variability and change are currently missing from the literature for time series with seasonality, although the scientific interest in the more general topic of investigating hydroclimatic features at the seasonal and monthly time scales at different parts of the world (see, e.g., the analyses by Nigam and Ruiz-Barradas 2006; Ljungqvist et al. 2016; PAGES Hydro2k Consortium 2017; Thomas and Nigam 2018; Papacharalampous et al. 2021, and the review by Koster et al. 2017) is undoubtedly large.

Herein, we fill this specific knowledge gap. More precisely, our aims are to:

(i) Devise a methodological framework for (a) the comprehensive characterization of the Earth's seasonal hydroclimate and its dynamics, and (b) the improved understanding of climate classification systems in terms of their seasonal hydroclimatic properties, through detailed feature investigations and comparisons across climate classes and continental-scale regions.

(ii) Apply the new framework to global temperature, precipitation and river flow time series datasets, thereby providing the first scientific insights on the targeted topic.

The remainder of the paper is structured as follows: Section 2.1 presents information on the investigated global datasets and their attentive use towards reaching our second aim, while Section 2.2 covers the feature estimation methodology, and the methodology followed for comparing the features across climates and continental-scale regions. Furthermore, Section 2.3 describes some additional methodological steps taken for ranking the investigated features according to how indicative they are in explaining roughly summarized climate or geographical information. These latter methodological steps are based on explainable machine learning, the role of which in delivering scientific insights and discoveries in natural sciences has been extensively discussed by Roscher et al. (2020). The results are presented in Sections 3.1, 3.2, 3.3 and 3.4, which are devoted to the global summaries, the summaries across climates, the summaries across continents and the comparisons enabled by explainable machine learning, respectively. The most important findings are further elaborated and discussed with respect to their theoretical and practical implications, as well as their links with the climate literature, in Section 4. In the same section, the strengths and limitations of the work are outlined. The paper concludes with Section 5, where the findings are summarized in the form of take-home messages.

## 2. Data and methods

### 2.1 Seasonal hydroclimatic time series, geographical divisions and climate classes

Peterson and Vose (1997), Do et al. (2018) and Menne et al. (2018) have compiled and made publicly available high-quality global time series datasets that summarize earth-observed hydroclimatic quantities. Starting from 13 104 mean monthly temperature, total monthly precipitation and mean monthly river flow time series that (i) originate from the aforementioned open time series datasets (see the information on their availability in Appendix A) and (ii) satisfy a series of specific length and quality conditions, we compute 13 104 39-year-long quarterly time series of 3-month means. These latter time series are the ones analysed herein in terms of their features (according to Section 2.2), after being standardized, and are referred to hereafter as "seasonal" time series. Each of them starts from winter and ends to fall, with winter, spring, summer and fall being represented by the month sets {December, January, February}, {March, April, May}, {June, July, August} and {September, October, November}, respectively. This specific seasonal representation scheme is adopted both in hydrology and hydroclimatology for conducting investigations at the global scale (see, e.g., Angell 1988; Dai and Wigley 2000; Arnal et al. 2018); therefore, it is considered appropriate for reaching the aims of the present work. Also notably, the length and quality conditions imposed for time series selection are the same with those outlined in the data sections of the earlier works by Papacharalampous et al. (2021, 2022). In summary, all the monthly hydroclimatic time series of these earlier works (that are used to compute the seasonal hydroclimatic time series of this work) had to be complete and with the same length, and a good compromise between their number and their length had to be made. Indeed, the available longer complete time series are much fewer. As regards the quality conditions, the monthly temperature and precipitation time series did not require additional testing other than that already applied during their original formation (Peterson and Vose, 1997; Menne et al., 2018). On the other hand, the monthly river flow time series had to be filtered through visual quality inspections for identifying irregularities (e.g., abrupt changes in their mean and variance) that could be due to human activities (Papacharalampous et al. 2021).

In greater detail, the herein analysed temperature, precipitation and river flow time series originate from 2 432, 5 071 and 5 601 geographical locations, respectively. These geographical locations are presented in Figure 1. In the same figure, climate classification information for these geographical locations is provided in terms of percentages both at the global and at continental scales, and regional groups of stations that are characterized by large or medium densities of observational stations are defined. The climate classification information is based on the well-established and interpretable system by Kottek et al. (2006), an updated version of the Köppen-Geiger climate classification system. The latter is the first quantitative system of its kind, as well as the most frequently used one (Kottek et al. 2006; Belda et al. 2014). It has been introduced by Wladimir Köppen and has later been made available as a world map updated by Rudolf Geiger (see Köppen 1936), while other historical information on its formation can be found in Thornthwaite (1943). The exact numbers of temperature, precipitation and river flow stations representing the various climate classes in our global hydroclimatic time series datasets (see also Kottek et al. 2006 for the statistical criteria underlying their definition) are presented in Figure S1 of the supplementary material (see Appendix B), while the respective counts of stations representing the main climate divisions (else referred to in the literature and in what follows as "climate zones") can be found in Figure S2 of the same material. These climate zones are the equatorial (A), arid (B), warm temperate (C), snow (D) and polar (E) ones, and are defined by specific temperature or precipitation conditions that allow the growth of different vegetation groups.

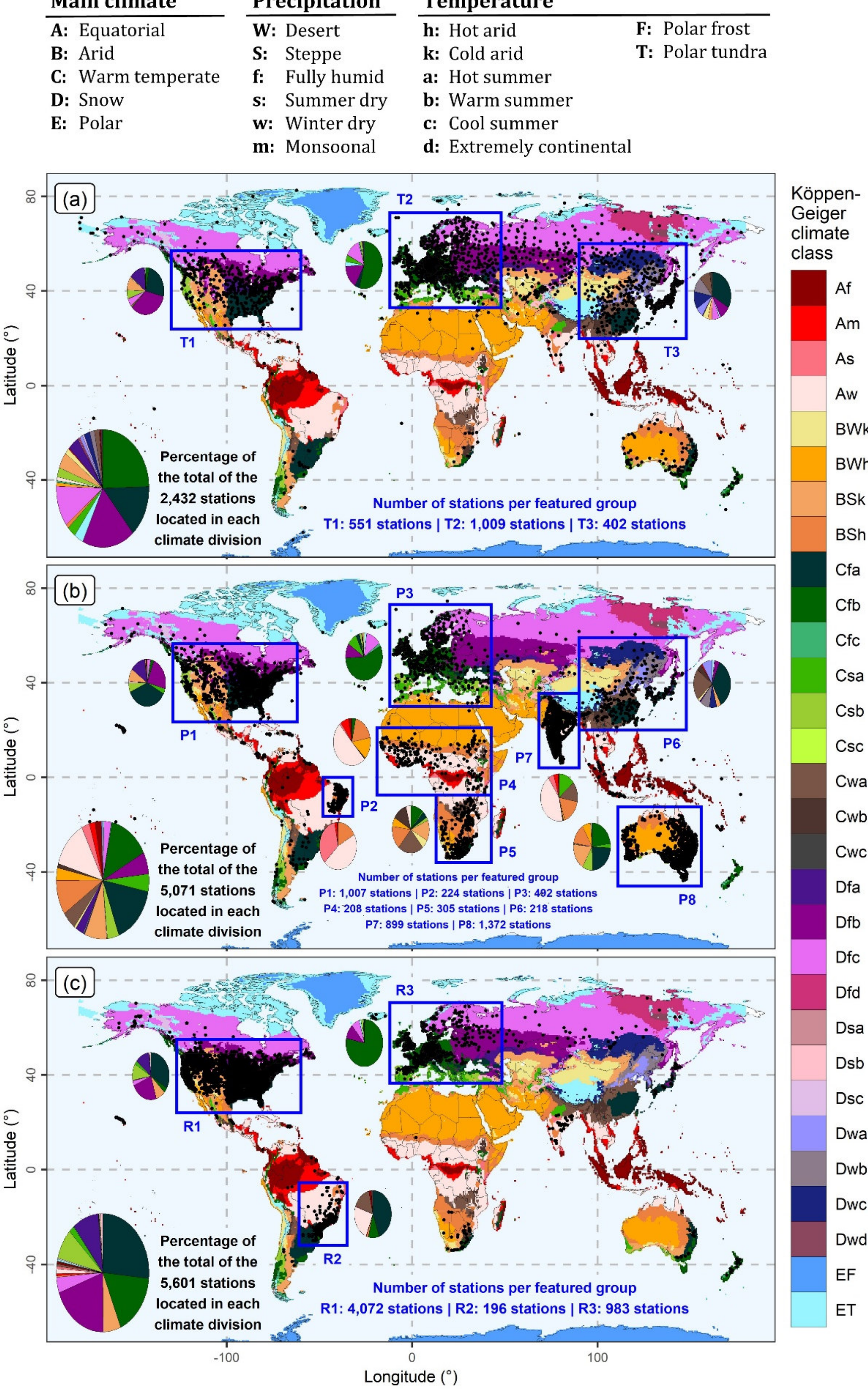

Figure 1. Geographical locations of the (a) temperature, (b) precipitation and (c) river flow stations exploited in the present work, climate classification information following the system by Kottek et al. (2006) and definition of regional groups of stations.

### 2.2 Feature computation, summaries and comparisons

We separately characterize each quarterly hydroclimatic time series (see Section 2.1) by estimating eight interpretable features (see Table 1). Three of these features belong to the large family of the sample autocorrelation features (see, e.g., Wei 2006), thereby being appropriate for investigating the temporal dependence structure. The first of them is the lag-1 sample autocorrelation of the time series (hereafter referred to simply as "lag-1 autocorrelation"). This feature characterizes both the magnitude and the sign of the correlation between two subsequent points in the time series and, thus, between the hydroclimatic means corresponding to two subsequent seasons for the analyses herein. The second autocorrelation feature is the sum of the squared sample autocorrelation values of the time series at the lags equal from 1 to 10 seasons. This specific feature is hereafter referred to simply as "autocorrelation summary", as it summarizes multiple autocorrelation features. The third autocorrelation feature supporting the investigations of this work is the lag-4 sample autocorrelation of a time series, a feature particularly relevant to characterizing the magnitude of the relationship between time series values corresponding to the same season (i.e., the annual-scale dependence) in a quarterly time series and, thus, hereafter referred to simply as "seasonal autocorrelation".

Table 1. Features computed for characterizing each quarterly hydroclimatic time series.

| S/n | Name | Brief description |
|---|---|---|
| 1 | Lag-1 autocorrelation | Lag-1 sample autocorrelation of the time series |
| 2 | Autocorrelation summary | Sum of the squared sample autocorrelation values of the time series at the lags equal from 1 to 10 seasons |
| 3 | Seasonal autocorrelation | Lag-4 sample autocorrelation of the time series |
| 4 | Temporal variation | Standard deviation of the first-order differenced standardized time series |
| 5 | Spectral entropy | Spectral entropy of the time series estimated by applying the method by Jung and Gibson (2006) |
| 6 | Hurst exponent | Hurst parameter of the fractional Gaussian noise process estimated by applying the method by Tyralis and Koutsoyiannis (2011) on the deseasonalized time series |
| 7 | Trend strength | Strength of trends in the time series estimated according to Kang et al. (2017) |
| 8 | Seasonality strength | Strength of seasonality in the time series estimated according to Kang et al. (2017) |

Another feature supporting our investigations herein is called "temporal variation". This feature is the standard deviation of the first-order differenced standardized time series and is similar to the coefficient of variation. For detailed information on time series differencing and its usefulness in reducing seasonality and trend features before conducting further analyses, the reader is referred to Hyndman and Athanasopoulos (2021). Moreover, we compute the "spectral entropy" of the time series by following the method by Jung and Gibson (2006). This feature is usually perceived to be measuring the random (or noisy) component of the time series (see, e.g., Hyndman and Athanasopoulos 2021) and, therefore, it facilitates characterizations and comparisons in terms of predictability. The sixth feature investigated in this work is the Hurst parameter of the fractional Gaussian noise process, an analogous of the autoregressive fractionally integrated moving average process (see, e.g., Wei 2006). This feature is computed by applying the method by Tyralis and Koutsoyiannis (2011) to time series that have been previously deseasonalized according to the additive classical model for time series decomposition (see, e.g., Hyndman and Athanasopoulos 2021) and by assuming four seasons. It is herein referred to as "Hurst exponent" and is broadly known to support long-range temporal dependence (or "Hurst phenomenon") investigations. While the term "autocorrelation feature" could also be used to refer to Hurst features, in what follows we will be using it to refer only to the lag-1 autocorrelation, autocorrelation summary and seasonal autocorrelation features.

The computation of the last two time series features of this work involves seasonal and trend time series decomposition using Loess (STL decomposition; see, e.g., Hyndman and Athanasopoulos 2021, Chapter 3.6). The exact procedures followed for obtaining the seasonal, smoothed trend and remainder components of each time series can be found in Hyndman and Khandakar (2008). Once these procedures have been completed, the "trend strength" and "seasonality strength" of the time series are computed according to Kang et al. (2017). Then, the former of these features is equal to 1 minus the quotient of the remainder component's variance and the variance of the deseasonalized time series, while the latter of them is equal to 1 minus the quotient of the remainder component's variance and the variance of the time series with its smoothed trend component removed.

We characterize the Earth's seasonal hydroclimate by providing global summaries, in the forms of histograms and means, of the previously computed features (see above) of the seasonal temperature, precipitation and river flow time series. Most importantly, we provide feature summaries and comparisons across climates and continents. The latter investigations are enabled by side-be-side boxplots and mean value computation and are conducted conditional upon the information summarized in Figure 1. More precisely, we summarize the results per climate class, per main climate division and per featured group of stations, with each of these groups representing a different pair {time series type, geographical division}. While all the geographical locations are taken into consideration in the computation of the global and regional summaries, the climate classes and main climate divisions represented by less than 30 stations are not studied and compared with the remaining ones for ensuring the sufficient representativeness of our results.

## 2.3 Feature importance comparisons

We apply explainable machine learning to compare the eight time series features of this work (see Section 2.2) with respect to their relevance as explanatory variables in two types of classification settings that differ in their dependent variables. More precisely, the investigated classification settings target at predicting −in the form of a best guess− and explaining-interpreting either (i) the main climate division from which a seasonal hydroclimatic time series originates or (ii) the group of stations and, therefore, the geographical division from which a seasonal hydroclimatic time series originates (only for those time series originating from one of the groups of stations featured in this work; see Figure 1), given only the values of the eight features characterizing this time series. These classification settings allow objective comparisons of the magnitude of the relationships between the seasonal hydroclimatic features assessed in this work and other Earth's features, and are studied separately for seasonal temperature, precipitation and river flow; therefore, we study 2 (number of classification setting types) × 3 (number of time series types) = 6 classification problems.

Each time, we fit random forests (Breiman 2001) for classification with 500 trees and compute two variable importance measures, namely the "mean decrease accuracy" and "mean decrease Gini" ones. We do not optimize the parameters of random forests, as according to the literature (see, e.g., the review by Tyralis et al. 2019): (a) the performance of this algorithm is expected to increase with increasing the number of trees, and (b) the default values of its remaining parameters have been empirically proven to be adequate. In summary, the first variable importance measure is computed as follows (Liaw 2018): For each tree, the error rate on the out-of-bag portion of the data is computed. The computation is repeated after permuting each predictor variable. The difference between the two are then averaged over all trees, and normalized by the standard deviation of the differences. If the standard deviation of the differences is equal to 0 for a variable, the division is not done (but the average is almost always equal to 0 in that case). The second metric is the total decrease in node impurities from splitting on the variable, as measured by the Gini index and averaged over all the trees (Liaw 2018). Once all the variable importance scores have been computed, the time series features are ranked based on them. The rankings are made separately for each set {time series type, dependent variable, variable importance measure}. The variable importance scores are not compared in terms of their magnitude, as such comparisons have only limited reliability (Tyralis et al. 2019).

## 3. Results

### 3.1 Overall summary of the Earth's seasonal hydroclimate

The global summaries of the seasonal temperature, precipitation and river flow features are provided in Figure 2. As the strengths of these features vary largely from geographical location to geographical location with few exceptions (referring to specific temperature features; see Figure 2a, b, g, j, m, v), providing characterizations and comparisons across climates and continents becomes even more important from both a scientific and a practical point of view. Aside from this important remark, other interesting results also stem from Figure 2. For instance, the lag-1 autocorrelation of the seasonal temperature time series is very close to zero, indicating almost uncorrelated variables due to the remarkably strong seasonality patterns at the quarterly time scale. These patterns seem to be manifesting themselves into pairs of consecutive values in each time series that are much less correlated than in other time scales, such as the monthly one. Notably, the same is not observed for the precipitation, and especially for the river flow, time series. Also notably, the autocorrelation summary feature takes remarkably large values for seasonal temperature, while the same does not apply to the cases of seasonal precipitation and river flow. Indeed, while lag-1 autocorrelation is almost zero for seasonal temperature and has absolute values that are much larger than zero for seasonal precipitation and river flow, it seems that the autocorrelation values at other lags (among those equal from 2 to 10 seasons) are much larger for seasonal temperature than for seasonal precipitation and river flow.

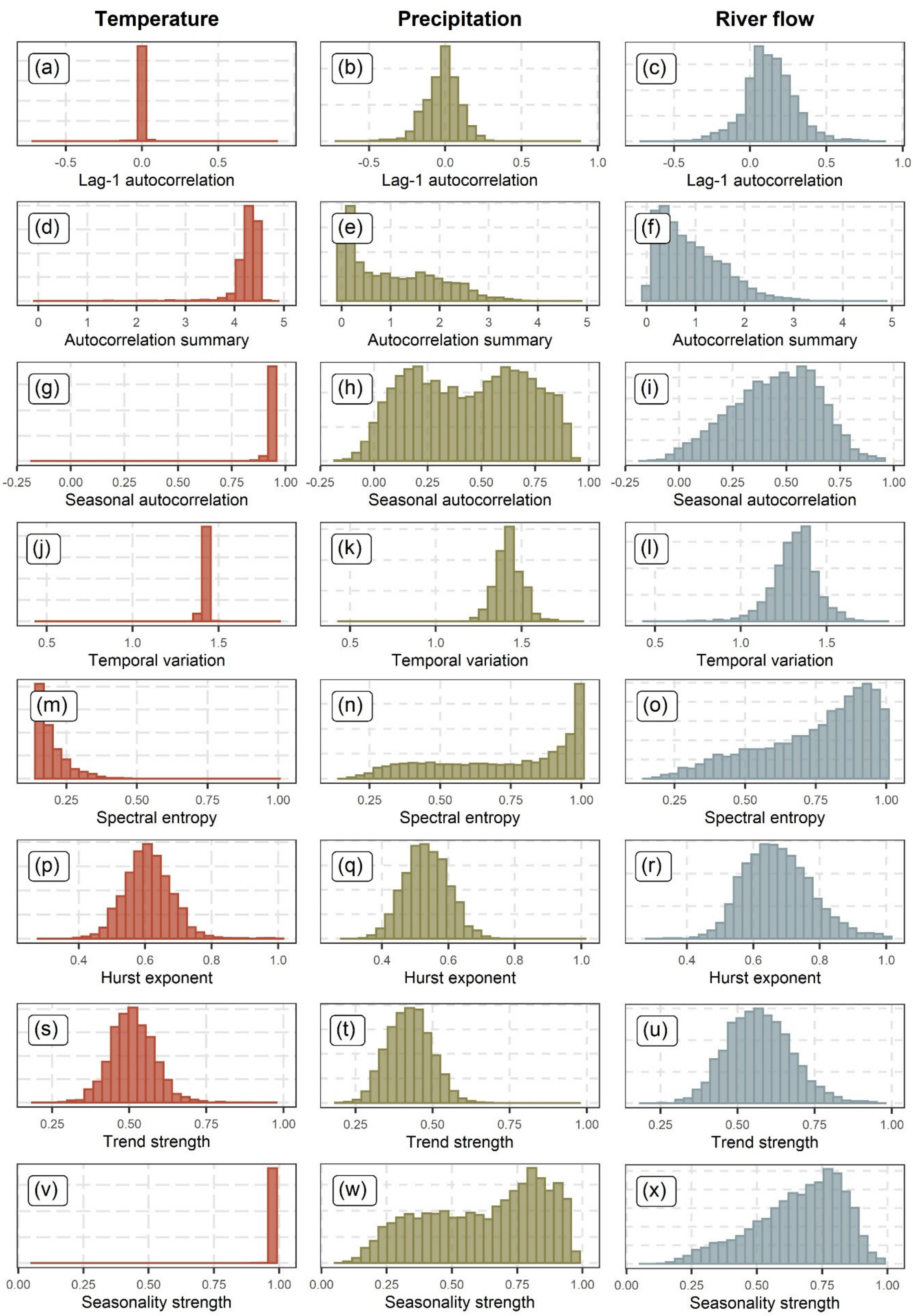

Figure 2. Global summaries of the seasonal temperature (a, d, g, j, m, p, s, v), precipitation (b, e, h, k, n, q, t, w) and river flow (c, f, i, l, o, r, u, x) features. For comparison purposes, the limits of the horizontal axes have been set common for the histograms in the same row. The relationships between the features can be inspected through Figures S3–S5 of the supplementary material.

### 3.2 Seasonal hydroclimatic feature comparisons across climates

Statistical summaries of the features of the seasonal temperature, precipitation and river flow time series across the investigated Köppen-Geiger climate classes are presented in Figures 3–5, while the respective summaries per main climate division are presented in Figures 6 and 7. These summaries offer a new basis for comparing the various Köppen-Geiger climates, other than the one offered by their original definitions (see the related discussions in Section 4). Moreover, especially Figures 5 and 7 offer information that could be exploited for constraining the uncertainty in the design of seasonal stochastic models. This latter contribution holds particularly for geographical locations with short or no time series records. In what follows, Figures 3–7 are collectively summarized and discussed, as they are interconnected.

In brief, both the Hurst phenomenon and trends are here found to be more intense for the seasonal temperature time series originating from the equatorial zone (A) than for those originating from other Köppen-Geiger climate zones (Figures 6f, g and 7a). This climate zone has minimum monthly temperatures equal or less than 18 °C (Kottek et al. 2006). Notably, seasonal temperature time series from this zone are further found to exhibit comparably large and even larger trends and long-range dependence, on average, than the seasonal river flow time series (Figures 6f, g and 7a, c). The same also holds (but to a less extent) for the seasonal temperature time series from the polar zone (E; composed by climates with maximum monthly temperatures less than 10 °C; Kottek et al. 2006). The polar zone is mostly represented in the examined temperature dataset by polar tundra climates (ET) and is here found to exhibit more intense seasonal temperature trends and long-range dependence compared to the remaining main climate divisions (i.e., the arid, warm temperate, and snow ones; Figures 6f, g and 7a, c).

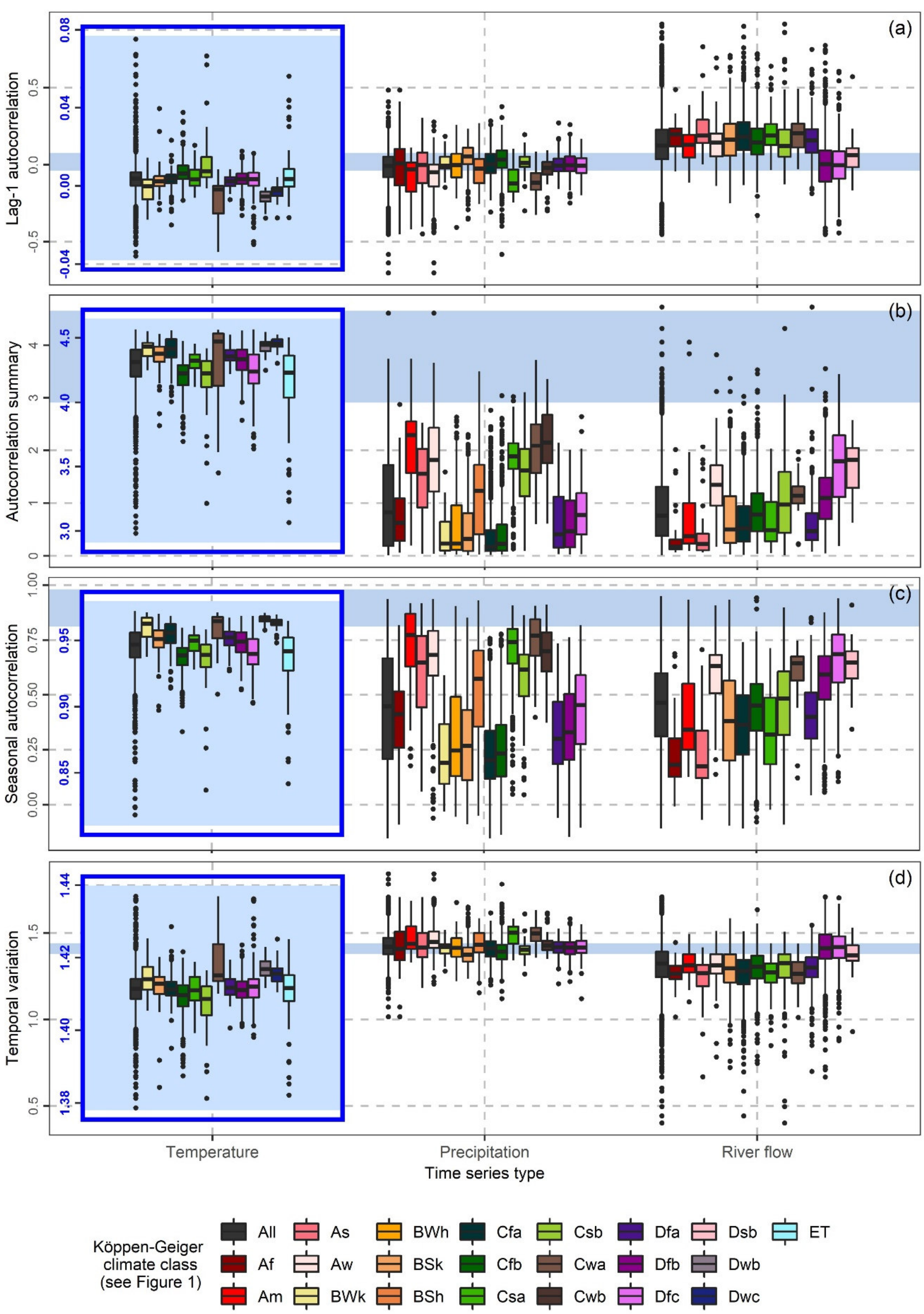

Figure 3. Side-by-side boxplots of the values of (a) lag-1 autocorrelation, (b) autocorrelation summary, (c) seasonal autocorrelation and (d) temporal variation per climate division (Kottek et al. 2006) for seasonal temperature, precipitation and river flow. The limits of the vertical axes of the blue rectangles have been set identical to the upper and lower limits of the blue ribbons, thereby allowing comparisons both within and across time series types. The corresponding mean values are presented in Figure 5.

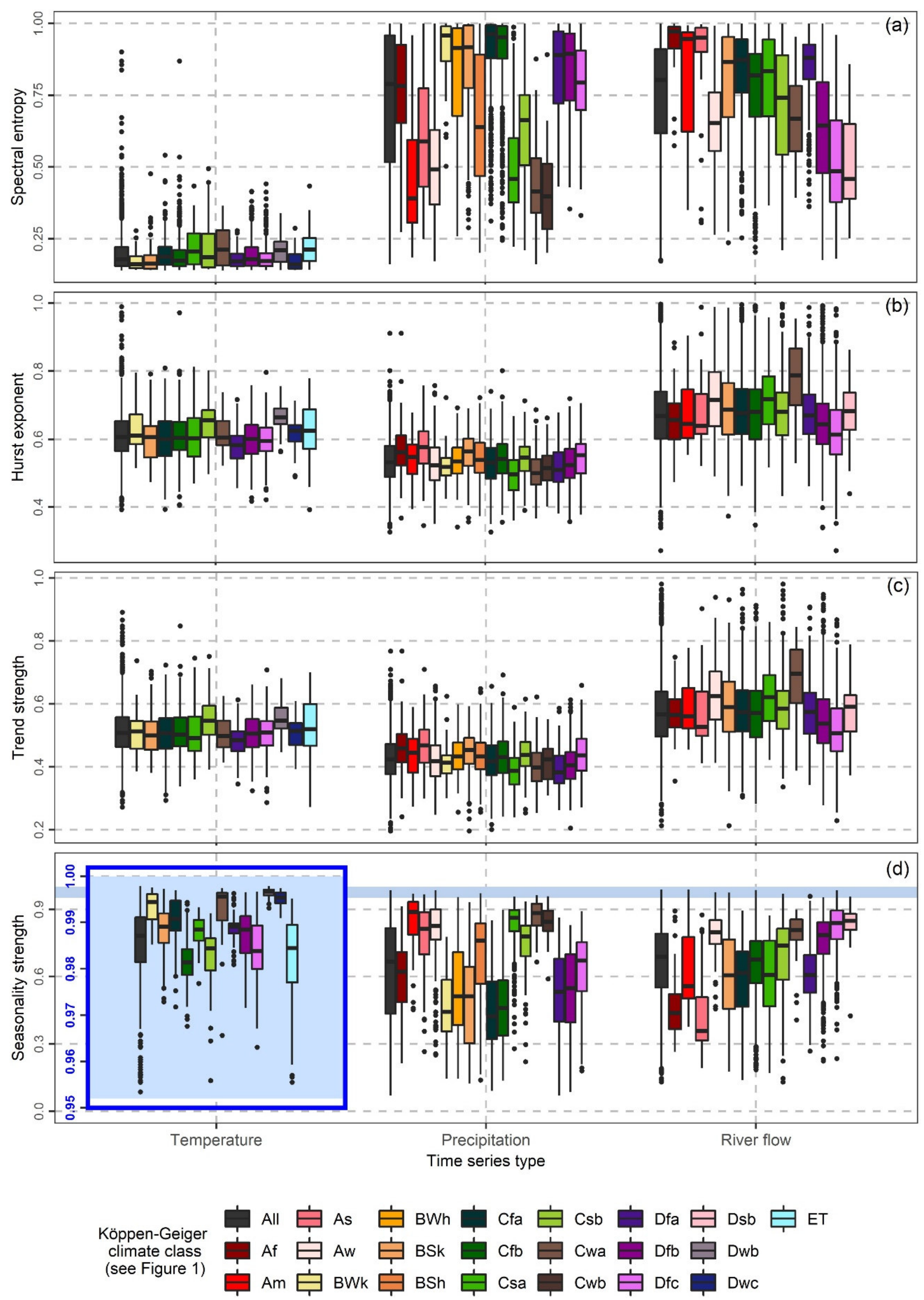

Figure 4. Side-by-side boxplots of the values of (a) spectral entropy, (b) Hurst exponent, (c) trend strength and (d) seasonality strength per climate division (Kottek et al. 2006) for seasonal temperature, precipitation and river flow. The limits of the vertical axis of the blue rectangle have been set identical to the upper and lower limits of the blue ribbon, thereby allowing comparisons both across the various temperature time series and across time series types. The corresponding mean values are presented in Figure 5.

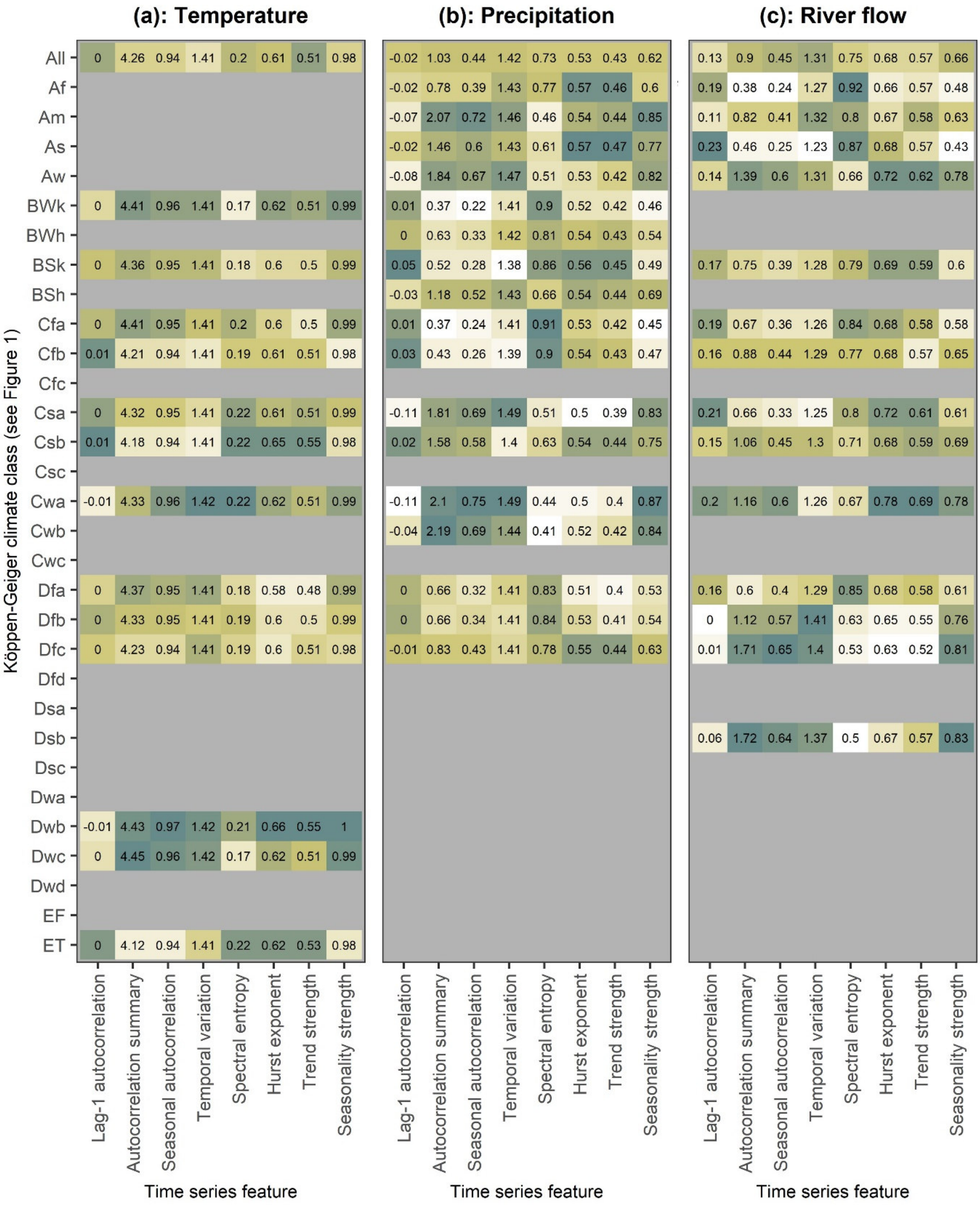

Figure 5. Mean values of the seasonal (a) temperature, (b) precipitation and (c) river flow features in the various climate classes (Kottek et al. 2006). The mean values presented in each column have been ranked and coloured from the smallest (white) to the largest (dark green), thereby highlighting even the smallest differences. The grey areas (i.e., areas lacking statistical summaries) correspond to underrepresented climate classes in the investigated global datasets (see Figure S1 in the supplementary material). The corresponding boxplots are presented in Figures 3 and 4.

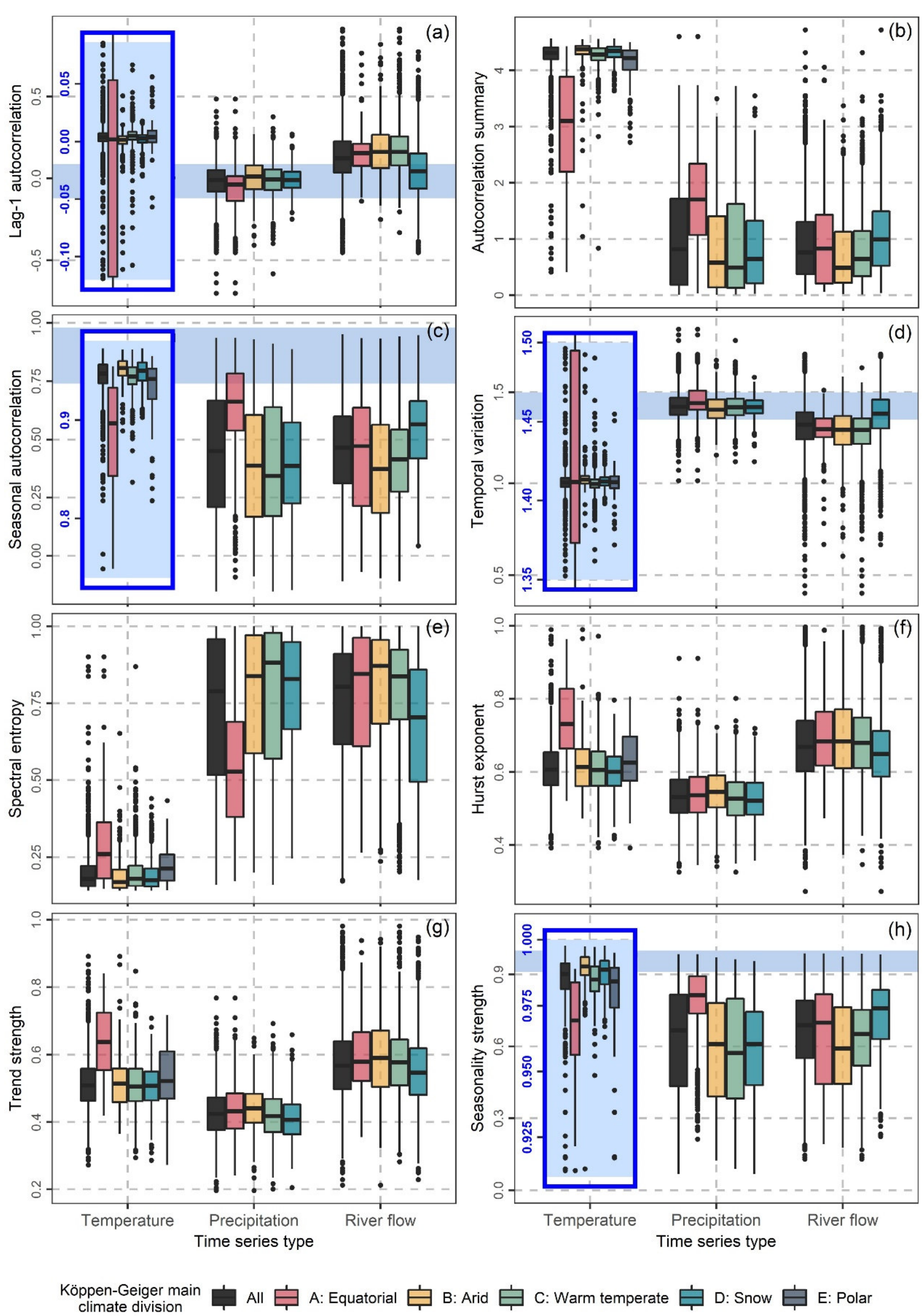

Figure 6. Side-by-side boxplots of the values of (a) lag-1 autocorrelation, (b) autocorrelation summary, (c) seasonal autocorrelation, (d) temporal variation, (e) spectral entropy, (f) hurst exponent, (g) trend strength and (h) seasonality strength per main climate division (Kottek et al. 2006) for seasonal temperature, precipitation and river flow. The limits of the vertical axes of the blue rectangles have been set identical to the upper and lower limits of the blue ribbons, thereby allowing comparisons both within and across time series types. The corresponding mean values are presented in Figure 7.

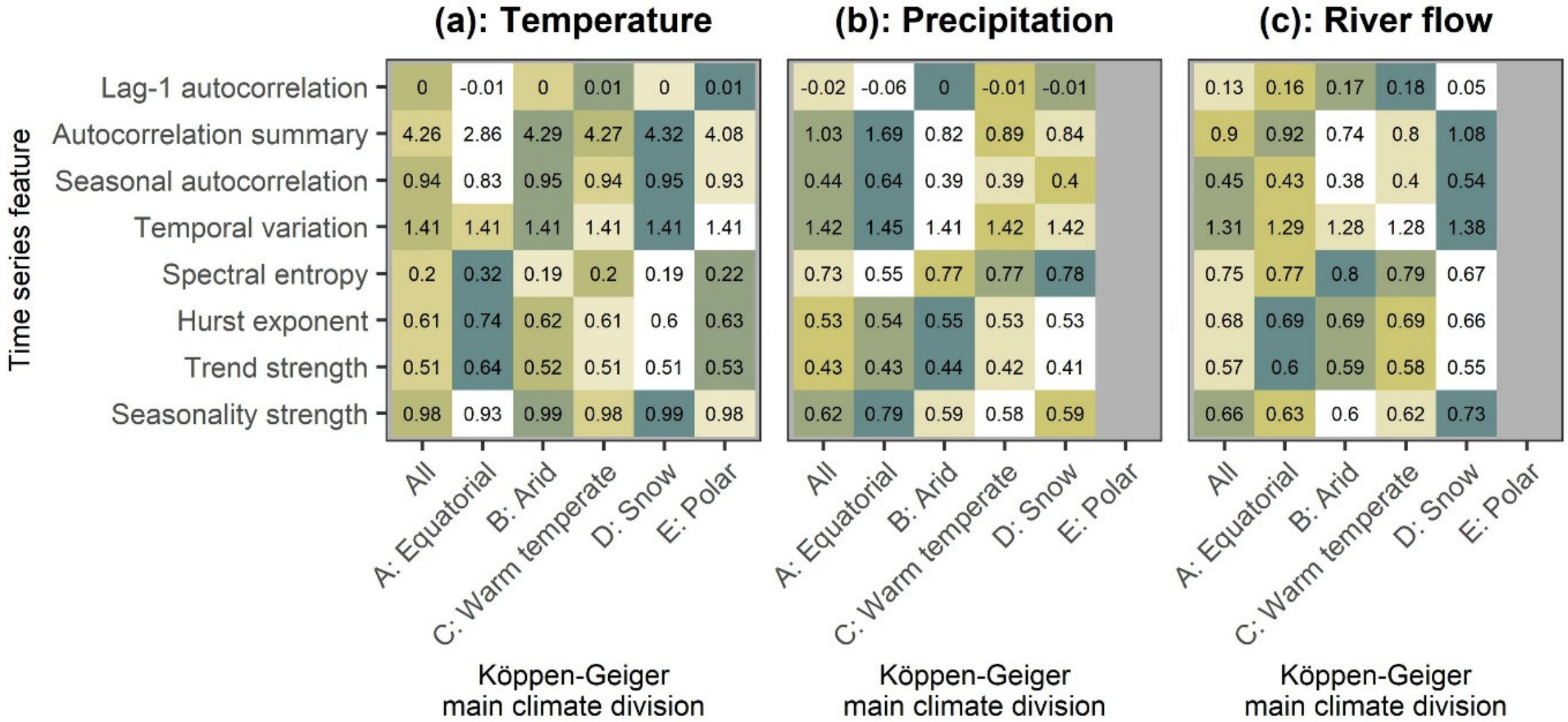

Figure 7. Mean values of the seasonal (a) temperature, (b) precipitation and (c) river flow features in the main climate divisions (Kottek et al. 2006). The mean values presented in each row have been ranked (separately for temperature, precipitation and streamflow) and coloured from the smallest (white) to the largest (dark green), thereby highlighting even the smallest differences. Polar climates are underrepresented in the investigated precipitation and river flow datasets (see Figure S2 in the supplementary material); thus, no statistical summaries are available for them in (b) and (c). The corresponding boxplots are presented in Figure 6.

Still, the same does not hold for all the individual Köppen-Geiger climate classes (defined by the main climate and subsequent precipitation conditions). Indeed, snow and winter dry climates with warm summer (Dwb) are here found to have the largest trend strength and Hurst exponent values among all the examined individual Köppen-Geiger climate classes, including the polar tundra climates (ET) (Figures 4b, c and 5a). Comparably large trend strength and Hurst exponent values have been computed, on average, for the warm temperate and summer dry climates with warm summer (Csb; Figures 4b, c and 5a). On the other hand, the least pronounced seasonal temperature trends and long-range dependence are found to characterize, again on average, the snow and fully humid climates with either hot (Dfa), warm (Dfb) or cool (Dfc) summer (Figures 4b, c and 5a). Apart from being characterized by the largest trends and long-range dependence, the seasonal temperature time series originating from equatorial climates have also the largest spectral entropy, followed by those originating from polar climates (Figures 6e and 7a). The latter climates are additionally found to have comparably large spectral entropy with three snow climate classes (specifically, the Cwa, Csa and Csb ones).

Links between seasonal temperature features with smaller interquartile ranges of values and the features defining the various Köppen-Geiger climates (Kottek et al. 2006) also emerge. For instance, it is found that the seasonal temperature time series originating from warm temperate and summer dry climates with warm summer (Csb) are those with the largest positive lag-1 autocorrelation (i.e., correlation between the mean values of two subsequent seasons in our investigations) and the smallest autocorrelation summaries, seasonal autocorrelation, temporal variation and seasonality strength, together with those originating from warm temperate and fully humid climates with warm summer (Cfb) and those originating from polar tundra climates (ET) (Figures 3a, b, c, d, 4d and 5a). It is additionally found that the seasonal temperature time series from warm temperate and winter dry climates with hot summer (Cwa) have the smallest negative lag-1 autocorrelation, followed by those from the snow and winter dry climates with either warm (Dwb) or cool (Dwc) summer (Figures 3a and 5a). These latter two climate types exhibit the most intense summary autocorrelation and seasonality features as regards their seasonal temperature time series. Furthermore, they exhibit equally pronounced temporal variation features with the Cwa climates (again as regards their seasonal temperature time series; Figures 3d and 5a).

The equatorial zone (A) also stands out because of its seasonal precipitation features. Indeed, the seasonal precipitation time series originating from this zone exhibit the largest autocorrelation summary, seasonal autocorrelation and seasonality strength values, and at the same time the smallest spectral entropy values, by far, compared to the seasonal precipitation time series originating from other climate zones (Figures 6b, c, h and 7b). Among the various climatic sub-regimes composing the equatorial zone, autocorrelation, temporal variation and seasonality features are more pronounced for the monsoonal ones (Am), which are also found to have equally intense autocorrelation and seasonality features with the warm temperate and winter dry climates with either hot (Cwa) or warm (Cwb) summer (Figures 3b, c, 4d and 5b). Again referring to the strength of the same time series characteristics, the Am, Cwa and Cwb climates are closely followed by the warm temperate and summer dry climates with either hot (Csa) or warm (Csb) summer and the equatorial climates with either dry summers (As) or winters (Aw) (Figures 3b, c, 4d and 5b). These latter four climate types are, in their turn, closely followed by the hot arid steppes (BSh), which also exhibit somewhat more intense autocorrelation and seasonality, on average, than the total of the examined seasonal precipitation time series, in contrast to the remaining climate classes (BWk, BWh, BSk, Cfa, Cfb, Dfa, Dfb), while those seasonal precipitation time series originating from snow and fully humid climates with cool summer (Dfc) are, on average, characterized by seasonality and summary autocorrelation features approximately equal to the global means (Figures 3b, c, 4d and 5b). Largely similar (but not identical) would be the ordering of the various climate classes regarding the temporal variation characteristics of the seasonal precipitation time series; however, in the case of these characteristics the magnitude differences are less pronounced (Figures 3d, 5b, 6d and 7b).

Similar to the summary autocorrelation and seasonality features of the seasonal precipitation time series, their entropy features are strongly related with the features defining the various Köppen-Geiger climates (Kottek et al. 2006), with the cold arid deserts (BWk), as well as the warm temperate and fully humid climates with hot (Cfa) or warm (Cfb) summer, being those with the largest mean values (Figures 4a and 5b). The opposite holds for three equatorial (Am, As, Aw) and four warm temperate (Csa, Csb, Cwa, Cwb) climates, while medium-magnitude entropy means characterize the seasonal precipitation time series of the remaining climates (Figures 4a and 5b). The trend and long-range dependence features of seasonal precipitation have smaller ranges of values compared to their entropy features; nonetheless, they are also somewhat related to the features defining the Köppen-Geiger climates (Figures 4b, c and 5b).

Mostly positive lag-1 autocorrelation is found to characterize the seasonal river flow time series originating from all the examined Köppen-Geiger climates apart from the snow and fully humid climates with warm (Dfb) or cool (Dfc) summer, and the snow and summer dry climates with warm summer (Dsb), for which the means (and the medians) of lag-1 autocorrelation of their seasonal river flow time series are quite close to (and, specifically, a bit larger than) zero (Figures 3a and 5c). The same three climate classes (Dfb, Dfc, Dsb) are found to also stand out for the pronounced autocorrelation summary, seasonal autocorrelation, temporal variation and seasonality strength features of their seasonal river flow time series, as well as for the smallest spectral entropy values of the same time series (Figures 3b, c, d, 4a, d and 5c), which also characterize the snow zone (D) in general (Figures 6b, c, d, e, h and 7c). This latter climate zone has minimum monthly temperatures less than −3 °C (Kottek et al. 2006).

Other Köppen-Geiger climates standing out for the pronounced autocorrelation summary, seasonal autocorrelation, temporal variation and seasonality strength features of their seasonal river flow time series are the equatorial savannahs with dry winter (Aw) and the warm temperate climates with dry winter and hot summer (Cwa), while the equatorial monsoonal climates (Am) and the equatorial savannahs with dry summer (As) are those with the least pronounced autocorrelation summary, seasonal autocorrelation, temporal variation and seasonality strength features (Figures 3b, c, d, 4d and 5c). These latter two climate classes are also characterized by the largest, among all the examined Köppen-Geiger ones, spectral entropy values of seasonal river flow time series, followed by the snow and fully humid climates with hot summer (Dfa), which also have among the smallest means of the autocorrelation summary, seasonal autocorrelation, temporal variation and seasonality strength features of their seasonal river flow time series (Figures 3b, c, d, 4a, d and 5c). Regarding the long-range dependence and trend features of the same time series, these are found to be more pronounced, by far, for the warm temperate climates with dry winter and hot summer (Cwa), with the equatorial savannahs with dry winter (Aw) and the warm temperate climates with dry winter and hot summer (Csa) following (Figures 4b, c and 5c). On the other hand, the seasonal time series from the snow and fully humid climates with cool summer (Dfc) are found to have the smallest trends and long-range dependence (Figures 4b, c and 5c).

### 3.3 Seasonal hydroclimatic feature comparisons across continents

Statistical summaries of the seasonal temperature, precipitation and river flow features across the various examined continental-scale regions (represented by specific groups of temperature, precipitation and river flow stations in our global hydroclimatic time series datasets; see Figure 1) are presented in Figures 8 and 9. These summaries could be used for reducing uncertainty in the design of seasonal stochastic models for geographical locations with short time series records, in a similar way to the previously discussed summaries. Their physical interpretation through Figures 1, 3, 4 and 5 is also possible to some extent. Rather small differences are found between seasonal temperature in North America, Europe and East Asia, with the latter continental-scale region having the most pronounced autocorrelation summary, seasonality, temporal variation, spectral entropy, long-range dependence and trend features of seasonal temperature time series, and Europe having the least pronounced autocorrelation summary, seasonality and temporal variation features (Figures 8 and 9a).

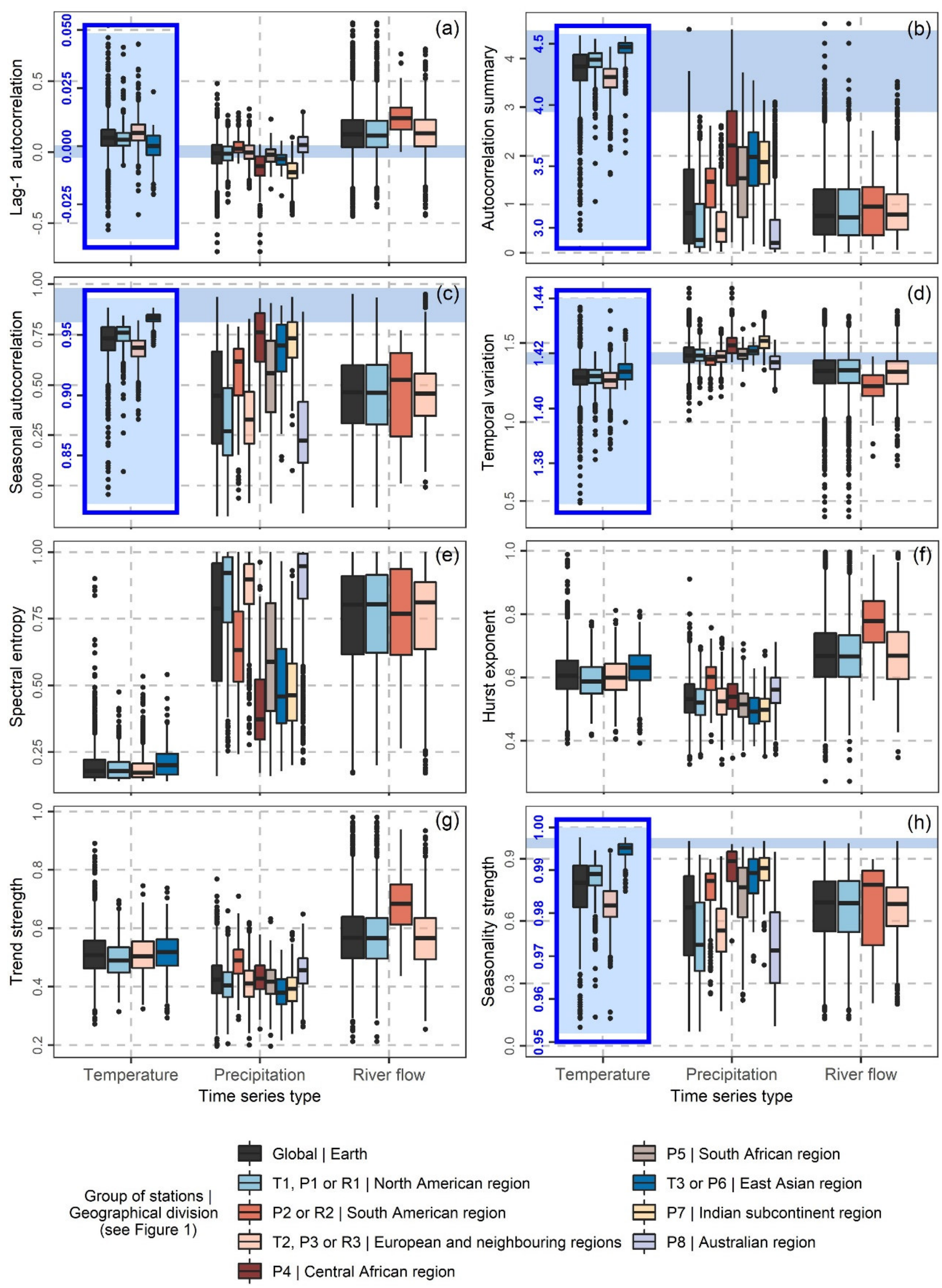

Figure 8. Side-by-side boxplots of the values of (a) lag-1 autocorrelation, (b) autocorrelation summary, (c) seasonal autocorrelation, (d) temporal variation, (e) spectral entropy, (f) Hurst exponent, (g) trend strength and (h) seasonality strength per featured group of stations (representing a specific geographical division) for seasonal temperature, precipitation and river flow. The limits of the vertical axes of the blue rectangles have been set identical to the upper and lower limits of the blue ribbons, thereby allowing comparisons both within and across time series types. The corresponding mean values are presented in Figure 9.

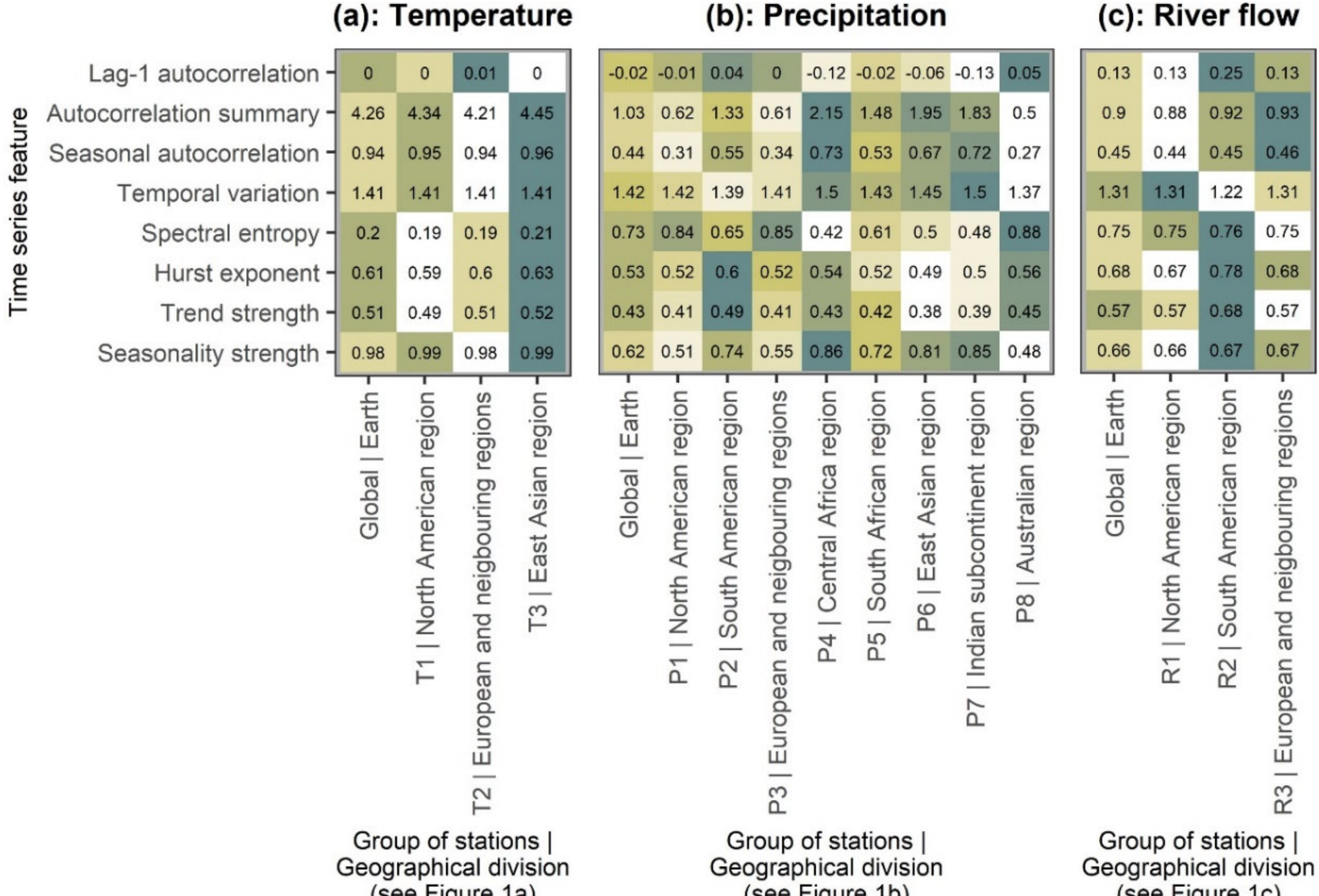

Figure 9. Mean values of the seasonal (a) temperature, (b) precipitation and (c) river flow features in the various featured groups of stations (representing different geographical divisions). The mean values presented in each row have been ranked (separately for temperature, precipitation and streamflow) and coloured from the smallest (white) to the largest (dark green), thereby highlighting even the smallest differences. The corresponding boxplots are presented in Figure 8.

The most notable differences from continent to continent are identified in terms of seasonal precipitation (Figures 8 and 9). In these terms, the most intense autocorrelation summary and seasonality features are found to characterize Central Africa. Two other continental-scale regions with comparably pronounced autocorrelation summary and seasonality features are East Asia and India, with the latter one having also the most intense temporal variation features of seasonal precipitation, followed closely by Central Africa and East Asia (Figures 8b, c, d, h and 9b). The same three geographical divisions also stand out for the spectral entropy values of their seasonal precipitation time series. These values are, on average, the smallest among those computed for the eight examined geographical divisions, while the largest are, by far, those characterizing North America, Europe and Australia (Figures 8e and 9b). Regarding the long-range dependence and trend features of the seasonal precipitation time series, these are somewhat more intense for South America than they are for the remaining examined geographical divisions, with Australia being next and this latter continent being followed, in its turn, by Central Africa (Figures 8f, g and 9b).

Among the three geographical divisions compared with respect to their seasonal river flow characteristics, South America is the one standing out. Indeed, the values of lag-1 autocorrelation, autocorrelation summary, seasonal autocorrelation, Hurst exponent, trend strength and seasonality strength of the time series originating from this latter geographical division are, on average, larger than those of the remaining two geographical divisions, while their temporal variation values are, on average, smaller (Figures 8 and 9c). Some similarities are additionally identified between the differences in the seasonal precipitation features of North America, South America and Europe, and the respective differences in the seasonal river flow features (Figure 8), probably because of the strong physical relationship between precipitation and river flow.

### 3.4 Feature importance in distinguishing climates and continents

Feature comparisons with respect to their importance-relevance in distinguishing the main climate divisions or the groups of stations (i.e., the T1, T2 and T3 ones for temperature, the P1, P2, P3, P4, P5, P6, P7 and P8 ones for precipitation, and the R1, R2 and R3 ones for river flow) are allowed by Figure 10. We observe that the spectral entropy, Hurst exponent and trend strength are ranked in the first three or four positions fewer times than the remaining features. Notably, differences in the feature values that are small in absolute terms (but significant in relative terms across climates and continents; see, e.g., the differences in the side-by-side boxplots in most of the blue rectangles of Figures 3, 4, 6 and 8) might be more relevant to the distinction between the climate zones or the groups of stations than differences that are larger in absolute terms (but less significant in relative terms).

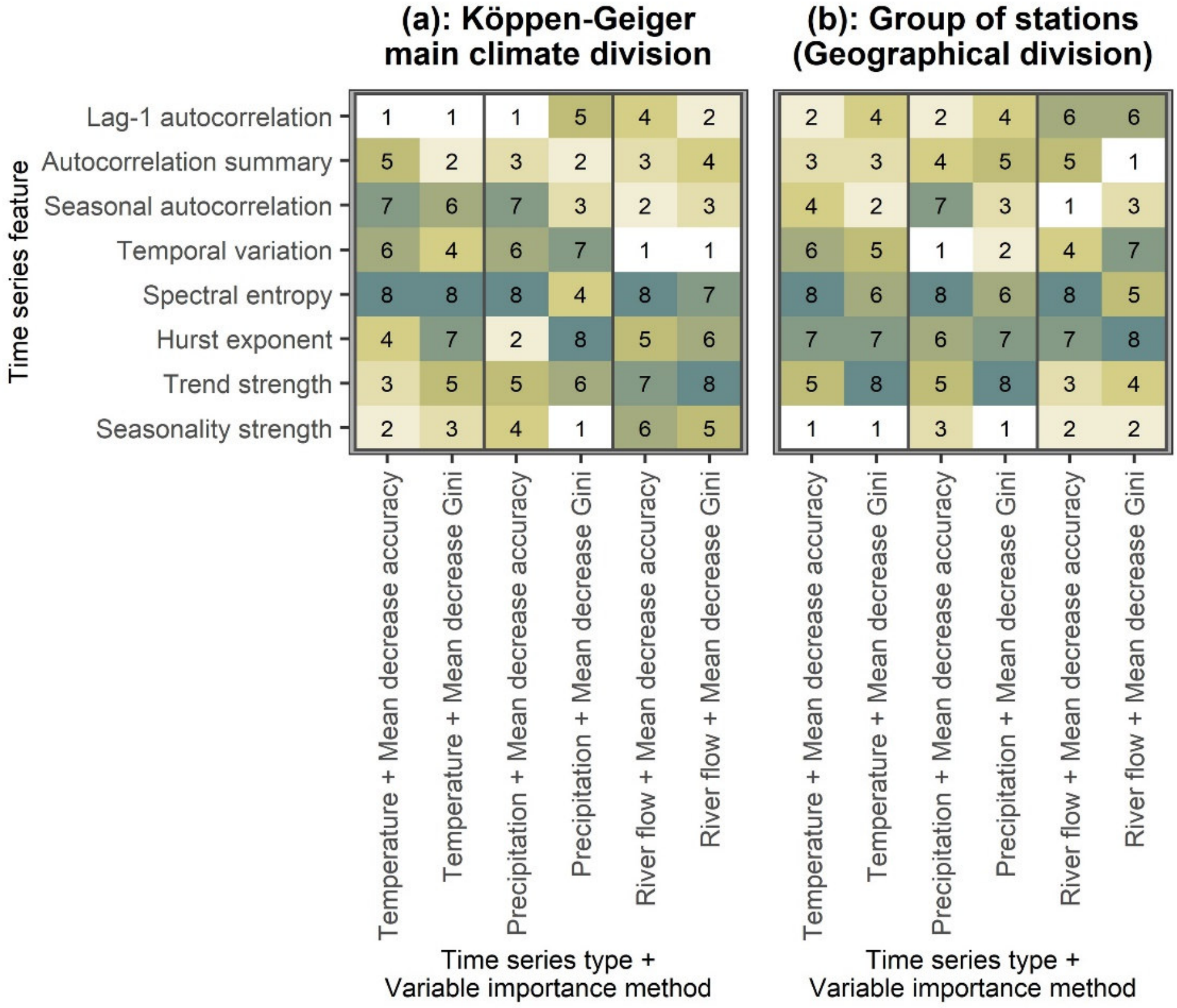

Figure 10. Rankings of the eight features from the most (1st) to the least (8th) important ones in distinguishing (a) the main climate divisions or (b) the groups of stations (i.e., the T1, T2, T3, P1, P2, P3, P4, P5, P6, P7, P8, R1, R2 or R3 ones) for temperature, precipitation and river flow using two variable importance methods.

## 4. Discussion

Our contribution to the literature rotates around eight interpretable time series features. These have been here computed, for the first time, for quantifying the differences in the Earth's seasonal hydroclimate across climates and continents, and even for questioning the existence of such notable differences. Facilitating direct comparisons in terms of temporal dependence, seasonality, temporal variation, entropy and trends at the global scale, our methodological framework has succeeded to address the aforementioned important targets, thereby substantially progressing our understanding and knowledge of the Earth's seasonal hydroclimate, in line with central themes, concepts and directions provided by the Panta Rhei initiative for the International Association of Hydrological Sciences (IAHS) scientific decade 2013–2022 (Montanari et al. 2013). Notable differences have been found in the strengths of most of the investigated temperature, precipitation and river flow features across the various Köppen-Geiger climate classes, as well as between several continental-scale regions. We, therefore, deem that (most of) the provided feature summaries across climates and continents (Figures 3−9) could lead to a larger reduction of the modelling uncertainties than the respective global summaries (Figure 2), and especially of those uncertainties that accompany the stochastic modelling and simulation of precipitation and river flow (but possibly of temperature as well) in areas with short or without earth-observed time series records. Reducing modelling uncertainties is traditionally among the most important targets in hydrology and hydroclimatology (Montanari et al. 2013; Blöschl et al. 2019). We also deem that the exploitation of our findings in this regard could be straightforward, given the direct utility of the computed statistics. For instance, these statistics could be used within feature-based time series simulation methodologies, such as the one proposed by Kang et al. (2020). In this particular case, each region or climate class could be represented by a set of features, with each of the latter taking simply as its value either the mean or the median of the values that have been computed for all the seasonal hydroclimatic time series representing this same region or climate class. Notably, the usefulness of this modelling approach depends on multiple factors, including the number of the time series available for each region or climate class and the spatial variability of the time series characteristics within each region or climate class.

From a different perspective, results of this work facilitate a better understanding of the various Köppen-Geiger climate classes. Indeed, these classes are here compared for the first time with respect to their seasonal hydroclimate in different terms than those followed for their original definition (see Kottek et al. 2006). Only a few of the relationships between the various hydroclimatic features defining these classes and the herein investigated hydroclimatic dependence, variability and change features could be expected. Yet, even for their case, additional quantitative and qualitative insights have emerged from our investigations. For instance, one could expect −already from existing climate knowledge− large seasonality strength to characterize the seasonal precipitation time series originating from equatorial monsoonal climates, a fact indeed confirmed and further quantified by our investigations. Nonetheless, one could not expect, among others, that seasonality strength of similar magnitude characterizes the seasonal precipitation time series originating from these latter climates and the warm temperate and winter dry climates with hot summer. Given the long history of the Köppen-Geiger climate classification system and its significance for the climate literature (see, e.g., the introduction sections in Kottek et al. 2006; Belda et al. 2014), we consider this aspect of our results particularly important. Of course, variants of the proposed methodological framework could also be applied in the future to better understand, from a seasonal hydroclimatic feature perspective, other climate classification systems (see, e.g., those by Thornthwaite 1948; Trewartha and Horn 1980; Fovell and Fovell 1993; Bunkers et al. 1996; Feddema 2005; Peel et al. 2007; Belda et al. 2014; Netzel and Stepinski 2016; Beck et al. 2018; Knoben et al. 2018; McCurley et al. 2021, and the historical landmark discussions on the formation of such systems by Thornthwaite 1943, 1948).

Overall, we consider the utilization of gauge-measured time series data as a strength of our methodological framework, given also the sufficiently large number of the investigated time series (i.e., 2 432 temperature, 5 071 precipitation and 5 601 river flow time series), their sufficient length (i.e., 156 values at the quarterly temporal scale) and their high quality. Alternatives would include the use of general circulation models or gridded (e.g., reanalysis) time series datasets, which offer optimal coverages of the Earth's surface and are not characterized by spatial heterogeneities. However, these models and datasets are known to contain large errors (see, e.g., the related discussions by Tyralis et al. 2018, as well as the related literature information provided therein) that could affect our hydroclimatic feature comparisons. As an inevitable consequence of the preference to gauge-measured data, a few Köppen-Geiger climate classes have been left out from our analyses. The same holds for continental-scale regions with low densities of data-rich stations. Moreover, comparisons between IPCC climate reference regions (Iturbide et al. 2020) would be another considerable way for extending the results of this work. Still, we believe to have provided extensive characterizations and comparisons leading to a broad overview of the Earth's seasonal hydroclimate.

## 5. Summary and conclusions

We have devised a new methodological framework for (a) the thorough characterization of the Earth's seasonal hydroclimate in terms of temporal dependence, seasonality, temporal variation, entropy and trends, and (b) the better understanding of climate classification systems in terms of their seasonal hydroclimatic properties. We have extensively applied the new framework by exploiting over 13 000 temperature, precipitation and river flow time series, in which the seasonal hydroclimatic behaviour is represented by 3-month means of earth-observed quantities. We have adopted the well-established and interpretable Köppen-Geiger climate classification system, and have characterized (most of) its classes based on their seasonal hydroclimatic feature values. We have further characterized continental-scale regions with large or medium density of observational stations, and applied explainable machine learning to compare the investigated feature types with respect to the amount of information that they provide for guessing either the main Köppen-Geiger climate or the continental-scale region. Our findings have both theoretical and practical implications, and are summarized with the following points, together with a few key strengths and limitations of the work:

- The Hurst phenomenon and trends are more intense for the seasonal temperature time series originating from the equatorial zone than for those originating from the remaining climate zones. Nonetheless, each of the various equatorial climate classes is under-represented in the herein examined temperature data, a limitation that has prohibited further investigations and insights into their temperature environments.
- Between a variety of well-represented climate classes (including several arid, warm temperate and snow ones, as well as the polar tundras), snow and winter dry climates with warm summer have the largest seasonal temperature long-range dependence and trends, followed by the warm temperate and summer dry climates with warm summer.
- The differences identified in terms of long-range dependence and trends across the various climate classes for the seasonal temperature time series could support the design of better stochastic models. However, the respective differences identified in terms of autocorrelation, temporal variation and seasonality are expected to be less informative in engineering contexts due to their rather small magnitude.
- Even in terms of long-range dependence and trends, seasonal temperature does not differ between North America, Europe and East Asia as largely as it does between some of the investigated climate classes. Still, for the latter continental-scale region, the Hurst phenomenon and trends are somewhat more pronounced.
- Moreover, the equatorial zone in general, and its monsoonal, summer dry and winter dry climates in particular, are characterized by the most pronounced autocorrelation and seasonality features of seasonal precipitation, as well as the least pronounced entropy features, together with the warm temperate and winter or summer dry climates with either hot or warm summer.

- Notable differences in seasonal precipitation are found between continental-scale regions as well, with by far the most pronounced autocorrelation, temporal variation and seasonality features, as well as the least pronounced entropy features, characterizing Central Africa, East Asia and India. The opposite holds for Australia, while the seasonal precipitation time series from this latter continental-scale region, as well as those from North America, stand out for their pronounced long-range dependence and trend features.
- The snow and fully humid climates with warm or cool summer, the snow and summer dry climates with warm summer, the equatorial savannahs with dry winter and the warm temperate climates with dry winter and hot summer are those with the most pronounced summary autocorrelation, temporal variation and seasonality features of their seasonal river flow time series, while the opposite holds for the equatorial monsoonal climates and the equatorial savannahs with dry summer.
- The warm temperate climates with dry winter and hot summer are by far those with the most intense Hurst phenomenon and trends. At the same time, the seasonal river flow time series from North America are characterized by the most pronounced long-range dependence and trends.
- These latter time series are further characterized by larger correlations of their subsequent seasons and smaller temporal variation than those originating from North America and Europe. The latter two continental-scale regions exhibit quite similar summaries of seasonal hydroclimatic features, in general.
- Entropy, long-range dependence and trend features are found to be (roughly) less indicative of the main Köppen-Geiger climate division or the continent in predictive modelling contexts than the remaining investigated feature types. It should be noted, however, that this matter requires additional investigations. Indeed, the coverage of the earth by ground-based stations with seasonal hydroclimatic time series that are long enough for our analyses is currently largely heterogeneous, with many classes being underrepresented in the formed classification settings. Furthermore, obtaining seasonal time series using different 3-month aggregation schemes (in which winter, spring, summer and fall are represented by different month sets) could lead to deviations in the feature importance in the classification problems.

- Aside from progressing our understanding of the Earth's seasonal hydroclimate and the Köppen-Geiger climate classification, the feature summaries across climates and continents provided by the present work could also help in reducing –as much as possible– modelling uncertainties, especially those that accompany the stochastic modelling of precipitation and river flow in areas with short or without time series records, and given the high quality and expected accuracy of the utilized global datasets. They could also be used as a metric for the evaluation of the performance of satellite- and reanalysis-based data products, as well as the performance of climate model products, deviating from conventional evaluations (global RMSE, correlations, etc.) in the sense that they benefit more from the newest data science concepts (e.g., those synthesized by Hyndman et al. 2022).

Overall, we believe that the authentic and comparative character of the proposed framework has been the key to providing fresh useful insights into the Earth's seasonal hydroclimate and its dynamics. We also deem that possible extensions of this framework would be beneficial from both a scientific and a practical point of view, given the high potential characterizing the newest data science advancements and the rare appearance of largely comparative hydroclimatic characterizations at the global scale.

**Funding information:** This work was carried out within the project "Investigation of Terrestrial HydrologicAl Cycle Acceleration (ITHACA)" funded by the Czech Science Foundation (Grant 22-33266M).

**Acknowledgments:** The authors are grateful to the Associate Editor and Reviewers for their constructive remarks.

**Author contributions:** GP and HT conceptualized the work and designed its experiments with input from YM, PM and MH. GP and HT performed the analyses and visualizations, and wrote the first draft, which was commented and enriched with new text, interpretations and discussions by YM, PM and MH.

**Conflict of interest:** The authors declare no conflict of interest.

## Appendix A Data availability

The gridded climate data (Kottek et al. 2006) can be retrieved through the `R` package `kgc` (Bryant et al. 2017), while the original hydrometeorological data (i.e., the mean and total monthly time series of varying lengths) can be retrieved through the following links: (a) https://www.ncdc.noaa.gov/ghcnm/v4.php (for temperature; Menne et al. 2018); (b) https://www.ncdc.noaa.gov/ghcnm/v2.php (for precipitation; Peterson and Vose 1997); and (c) https://doi.org/10.1594/PANGAEA.887477 (for river flow; Do et al. 2018).

## Appendix B Supplementary material

Supplementary material can be found in the online version and includes additional visualizations (Figures S1–S5).